\documentclass{Rinton-P10x7}
\def\be{\begin{equation}}
\def\ee{\end{equation}}

\begin{document}

\title{On Dynamical Adjustment Mechanisms for the Cosmological 
Constant
\footnotemark[1]
}
\renewcommand{\thefootnote}{\fnsymbol{footnote}}
\footnotetext[1]{
talk at PASCOS 2001, Chapel Hill, North Carolina, 10-15 April 2001}

\author{A. Hebecker}

\address{Theory Division, CERN, CH-1211 Geneva 23, Switzerland\\
E-mail: arthur.hebecker@cern.ch
}

\maketitle

\abstracts{After recalling why dynamical adjustment mechanisms represent a 
particularly attractive possibility for solving the cosmological constant
problem, we briefly discuss their intrinsic difficulties as summarized in
Weinberg's no-go theorem. We then comment on some problems of the recently
proposed `self-tuning' mechanism in 4+1 dimensions. Finally, we describe
an alternative approach which uses the time-evolution of the universe
to achieve a dynamical relaxation of the cosmological constant to zero.
}

We are used to describing particle physics and cosmology by an appropriate 
low-energy effective field theory. For energies presently accessible in
the laboratory, this effective field theory contains the quantum fields of 
the standard model together with classical gravity. In this framework, one
expects the vacuum fluctuations of the quantum fields to produce a
cosmological constant corresponding to an energy density $\sim
\mu^4$, where the cutoff $\mu$ is at least ${\cal O}$(TeV). This has to be 
compared with the experimental upper bound $\sim {\cal O}$(meV$^4$). It is 
hard to imagine that this discrepancy can be resolved by any high-energy 
symmetry argument, since numerous low-energy standard-model contributions 
to the vacuum energy, which depend on all the details of the low-energy 
field theory, exist. An example is provided by the gluon condensate of QCD, 
the value of which depends on the non-perturbative dynamics. Even small 
variations of this condensate, associated, e.g., with the light quark 
masses, are sufficient to exceed the experimental upper bound on the 
cosmological constant by a vast amount. It is difficult to see how a 
high-energy mechanism, based, e.g., on supersymmetry or string theory, 
would `know' enough about all these effective low-energy parameters to 
compensate their contribution with the high accuracy required. 

The idea of a dynamical adjustment of the cosmological constant to zero 
represents an attractive alternative possibility (see~\cite{wein} for a 
review). Ideally, such a mechanism would ensure that, given the sudden 
appearance of a new contribution to the vacuum energy density (e.g., 
through a phase transition in late cosmology), an equal opposite-sign 
contribution is created and exponential expansion or collapse are avoided. 

Unfortunately, as was explained in~\cite{wein}, the most straightforward 
attempts of constructing such a mechanism are bound to fail. More 
specifically, it was shown that one can not build an action for gravity, 
standard model and a finite number of scalar fields which would, without 
fine-tuning, be extremized by a space-time independent field configuration 
with zero curvature. A very simple, intuitive argument can be based on the 
action
\be
S=\int d^4x\,\sqrt{-g}\,\left(\frac{1}{2}M^2R+{\cal L}_{SM}+{\cal L}(\phi)
\right)\,,
\ee
which, after integrating out the standard model degrees of freedom and
restricting oneself to constant fields $\phi$, takes the form
\be
S=\int d^4x\,\sqrt{-g}\,\left(\frac{1}{2}M^2R-\Lambda_{SM}-V(\phi)\right)\,.
\ee
It is now clear that one can not design a potential $V(\phi)$ such that the 
minimum value of $\Lambda_{SM}+V(\phi)$ is zero for any value of 
$\Lambda_{SM}$. Furthermore, the situation is not improved by generalizing 
the action according to $\Lambda_{SM}\,\to\,\Lambda_{SM}f(\phi)$,
as would be the case in Brans-Dicke-like theories\footnote{
One has to require that $[\Lambda_{SM}f(\phi)+V(\phi)]'=0$ implies 
$\Lambda_{SM}f(\phi)+V(\phi)=0$. This leads to the condition $f(\phi)= 
\mbox{const.}\times V(\phi)$, which does not solve the original problem.
}. 
As discussed in more 
detail in~\cite{wein}, the reason for these difficulties lies in the 
fact that, for constant fields, the action depends on the metric only 
through an overall factor $\sqrt{-g}$. Thus, the problem is reduced to the 
requirement that the minimal value of a potential is zero. Given the 
additive contribution $\Lambda_{SM}$, this leads to fine-tuning. 

To design a working adjustment mechanism, one needs to relax some of the 
assumptions of the above no-go theorem. For example, the low energy 
effective theory of gravity could be quite different from Einstein's general 
relativity without conflict with experiment. This has been demonstrated 
very impressively in~\cite{rs}, where a non-compact extra dimension gives 
rise to a continuum of gravitational Kaluza-Klein modes without a mass gap 
(the Randall-Sundrum II scenario). Indeed, starting with~\cite{led}, many 
attempts have been made to construct an adjustment mechanism in this 
framework. 

The original model is based on gravity + a scalar field in 4+1 dimensions, 
with the standard model fields being localized on a 3+1 d boundary (with 
orbifold boundary conditions). After integrating out the standard model
degrees of freedom, the action is
\be
S=\int d^4xdy\sqrt{-g_5}\left(\frac{1}{2}M_5^3R-\frac{3}{2}(\partial\phi)^2
\right)-\int d^4x\sqrt{-g_4}\Lambda_{SM}\exp\left(\frac{2\phi}{M_5^{3/2}}
\right)\,.\label{ac}
\ee
The essential observation is that, for any value of $\Lambda_{SM}$, the 
equations of motion derived from this action have a static solution with 
vanishing brane curvature. In the bulk, this solution has a curvature
singularity at finite proper distance from the brane. This gives rise to
the hope that low-energy 4d gravity will result since the fifth dimension 
is effectively finite.

The above `self-tuning' scenario has already been criticized on a rather 
fundamental level in~\cite{nil}. Nevertheless, we want to describe an 
objection to~\cite{led} which, although close in spirit to~\cite{nil}, might 
be somewhat simpler and more direct. The point is that the field  
configuration of~\cite{led} does not extremize the action Eq.~(\ref{ac}). 
This is immediately clear since the only non-derivative coupling of $\phi$ 
appears in the factor multiplying $\Lambda_{SM}$. Thus, for nonzero 
$\Lambda_{SM}$, changing $\phi$ by an infinitesimal constant leads to a 
change of $S$ proportional to that constant.

All this is not in contradiction to the claim of~\cite{led} that their 
field configuration solves the equations of motion (everywhere 
outside the singularity). The reason is that, when varying an action on a 
finite interval (as is appropriate if a singularity is present), one gets,
in addition to the equations of motion, a boundary term, which has not 
been considered in~\cite{led}. Therefore it appears that one either has to 
give up the action principle or to allow for the possibility of a 
non-derivative coupling of $\phi$ to the physics at the singularity.
In the latter case, it seems likely that some form of fine-tuning will, 
after all, be required. 

In the remainder of this paper, we want to outline a different way of 
relaxing the assumptions of Weinberg's no-go theorem~\cite{rub,hw}. In this 
approach, one does not insist on a flat universe as an extremum of the 
action, but satisfies oneself with a cosmology that approaches the flat 
geometry asymptotically as the universe grows old. 

A possible mechanism for such a `time-dependent' adjustment of the 
cosmological constant has been suggested in Rubakov's scenario of a 
relaxation at inflation~\cite{rub}. Developing his ideas, we present a 
time-dependent adjustment mechanism for the cosmological constant that can 
be at work in a realistic, late Friedmann-Robertson-Walker 
universe~\cite{hw}. 

In our model, the energy density is dominated by non-standard-model dark 
matter together with a quintessence field $\varphi$. As in Rubakov's scenario,
the cosmological constant, characterized by a field $\chi$, rolls down a 
potential and approaches zero asymptotically. This is realized by a kinetic 
term for $\chi$ that diverges as $t^4$ at large $t$. The effective $t^4$ 
behaviour is realized with the help of $\varphi$. The asymptotic stability 
of this solution is ensured by the coupling to a Brans-Dicke field $\sigma$.

The action of our model can be decomposed according to 
\be
S=S_E+S_{SF}+S_{SM}\,,
\ee
where $S_E$ is the Einstein action, $S_{SF}$ the scalar field action, and 
$S_{SM}$ the standard model action, which is written in the form
\be
S_{SM}=S_{SM}[\psi,g_{\mu\nu},\chi]=\int d^4x\sqrt{-g}\,
{\cal L}_{SM}(\psi,g_{\mu\nu},\chi)\,.\label{ssm}
\ee
Here $\psi$ stands for all standard-model fields. The scalar $\chi$ is 
assumed to govern the effective UV-cutoffs of the different modes of 
$\psi$, thereby influencing the effective cosmological constant. Units are 
chosen such that $16\pi G_N=1$. 

Integrating out the fields $\psi$, one obtains (up to derivative terms)
\be
S_{SM}=\int d^4x\sqrt{g}\,V(\chi)\,.
\ee
Let the potential $V(\chi)$ have a zero, $V(\chi_0)=0$ with $\alpha=V'( 
\chi_0)$, and rename the field according to $\chi\,\to\,\chi_0+\chi$. Then 
the action near $\chi=0$ becomes 
\be
S_{SM}=\int d^4x\sqrt{g}\,\alpha\chi\,.
\ee
Due to this potential the field $\chi$ will decrease (for $\alpha>0$) during 
its cosmological evolution. It can be prevented from rolling through the 
zero by a diverging kinetic term~\cite{sa}. 

First, let the geometry be imposed on the system, i.e., assume a flat FRW 
universe with $H=(2/3)\,t^{-1}$. With a kinetic lagrangian 
\be
{\cal L}_{SF}=\frac{1}{2}\,\partial^\mu\chi\partial_\mu\chi\,t^4\,,
\ee
one finds a solution where $\chi=(\alpha/6)\, t^{-2}$, which provides an 
acceptable late cosmology.

Clearly, it requires fine tuning of the initial conditions to achieve 
the desired behavior $\chi\to 0$ as $t\to\infty$. However, this fine 
tuning can be avoided by adding a Brans-Dicke field $\sigma$ that `feels' 
the deviation of $\chi$ from zero and provides the appropriate `feedback' 
to the kinetic term so that $\chi$ reaches zero asymptotically independent 
of its initial value. 

The field $\sigma$ has a canonical kinetic term and it is coupled to 
${\cal L}_{SM}$ by the substitution $g_{\mu\nu}\to g_{\mu\nu}\sqrt{\sigma}$ 
in Eq.~(\ref{ssm}), 
\be
S_{SM}=S_{SM}[\psi,g_{\mu\nu}\sqrt{\sigma},\chi]=\int d^4x\,\sigma\sqrt{g}\, 
{\cal L}_{SM}(\psi,g_{\mu\nu}\sqrt{\sigma},\chi)\,.\label{fco}
\ee
Integrating out the fields $\psi$, one obtains 
\be
S_{SM}=\int d^4x\sqrt{g}\,\alpha\sigma\chi\label{ssmf}
\ee
near $\chi=0$. The scalar field lagrangian is now taken to be 
\be
{\cal L}_{SF}=\frac{1}{2}(\partial\chi)^2\sigma^2t^4+\frac{1}{2}
(\partial\sigma)^2-\beta\sigma t^{-2}\,.\label{lsf}
\ee
This gives rise to the asymptotic solution $\chi=\chi_0\,t^{-2}$ 
and $\sigma=\sigma_0=\mbox{const.}$ This solution is stable, i.e., one finds 
the same asymptotic behaviour for a range of initial conditions. The 
stability does not depend on the precise values of the parameters $\alpha$ 
and $\beta$.

What remains to be done is the replacement of the various explicit 
$t$-dependent functions by the dynamics of an appropriate field. This can 
be achieved using the simplest version of quintessence~\cite{qui} with an
exponential potential $V_Q(\varphi)=e^{-a\varphi}$. One finds the late-time 
behaviour $\varphi=(2/a)\ln t$. The explicit time dependence in 
Eq.~(\ref{lsf}) can now be replaced by a coupling to $\varphi$. Technically, 
this is realized by the substitution $t^2\to e^{a\varphi}$ in 
${\cal L}_{SF}$. 

The complete lagrangian, including the curvature term and the effective 
standard model action, Eq.~(\ref{ssmf}), reads
\be
{\cal L}=R+\frac{1}{2}(\partial\chi)^2\sigma^2\,e^{2a\varphi}+\frac{1}{2}
(\partial\sigma)^2+\frac{1}{2}(\partial\varphi)^2+\alpha\sigma\chi+(1-\beta
\sigma)\,e^{-a\varphi} \,.
\ee
We have checked numerically that the resulting late time solution, which
gives rise to an acceptable cosmology, is stable with respect to small 
variations of all parameters and initial conditions. The only problem is
the too strong coupling of the Brans-Dicke field to baryons. However, this 
can probably be avoided in a more carefully constructed model or by the 
ad-hoc introduction of a kinetic term for $\sigma$ that grows for large 
$t$.

\end{document}